# A Reproducible Workflow for Scraping, Structuring, and Segmenting Legacy Archaeological Artifact Images

Juan Palomeque-Gonzalez

[jfpalomeque.com](jfpalomeque.com) | [github.com/jfpalomeque](github.com/jfpalomeque)

## Abstract

This technical note presents a reproducible workflow for converting a legacy archaeological image collection into a structured and segmentation ready dataset. The case study focuses on the Lower Palaeolithic hand axe and biface collection curated by the Archaeology Data Service (ADS), a dataset that provides thousands of standardised photographs but no mechanism for bulk download or automated processing. To address this, two open source tools were developed: a web scraping script that retrieves all record pages, extracts associated metadata, and downloads the available images while respecting ADS Terms of Use and ethical scraping guidelines; and an image processing pipeline that renames files using UUIDs, generates binary masks and bounding boxes through classical computer vision, and stores all derived information in a COCO compatible Json file enriched with archaeological metadata. The original images are not redistributed, and only derived products such as masks, outlines, and annotations are shared. Together, these components provide a lightweight and reusable approach for transforming web based archaeological image collections into machine learning friendly formats, facilitating downstream analysis and contributing to more reproducible research practices in digital archaeology

## Introduction

Traditionally, archaeological research produces a large amount of diverse data, usually compiled and processed for researchers for specific projects or questions, including not only tables and text files, but maps, drawings and images of artifacts. In recent years, online journals, and multiple initiatives such as ADS (*Archaeology Data Service*, 2025) or Trove (*Trove.scot*, 2025) have provided easy access to some of that corpus of data (Richards, 2017). However, actually accessing, downloading or obtaining the files you need is often far less straightforward than it should be.

In this technical note, we will explore the challenges to access to all the data from the dataset *Lower Palaeolithic technology, raw material and population ecology (bifaces)*(Marshall *et al.*, 2007), stored in ADS, a possible solution to bulk download all the files and additional metadata, and the design of a reusable pipeline to process the images, segmenting the artifacts and storing all the data in a machine learning friendly format (COCO compatible Json format) (Lin *et al.*, 2014; *COCO - Common Objects in Context*, 2023), allowing the integration of the dataset details with new datasets without risks of file name crashing, and facilitating further analysis.

# Source Dataset & Licensing

This dataset is part of a project developed between 1999 and 2001, designed to analyse Lower Palaeolithic lithic tools (Acheulian bifaces: handaxes, cleavers and picks) from different sites located in Africa, Europe and the Near East, with a date between 1.5Myr to 300Kyr ago. It contains 10668 standardised digitised images of 3556 bifaces, as well as information on provenience, raw material and standard measurements.

## Accessing the Dataset

In order to access the records, a query needs to be submitted using the provided form (figure 1) ([https://archaeologydataservice.ac.uk/archives/view/bifaces/bf_query.cfm](https://archaeologydataservice.ac.uk/archives/view/bifaces/bf_query.cfm)). For accessing all the available records, you can input an extremely high value in some of the fields (like max weight = 999999999 g). That would return all the records in a list view (figure 2). By clicking on any of the results, you can see the complete record, with all the three images per artifact, and all the recorded details (figure 3). By clicking on the different images, you can access a high quality copy that can be downloaded locally. And example of an image can be found on figure 4

## Query
Please enter your search terms below press the query button to search the database.

Explanation of the search options

[Submit] [Reset]

| | |
|---|---|
| Id no. | |
| Site name | |
| Country | |
| Biface type | |
| Raw material | |
| Edge profile | |
| Condition | |
| Percentage of circumference worked | |
| Museum | |
| Finder | |
| Weight (g) *(min - max)* | - |
| Length (mm) *(min - max)* | - |
| Breadth (mm) *(min - max)* | - |
| Breadth/Length *(min - max)* | - |
| L1/Length *(min - max)* | - |
| BA/BB *(min - max)* | - |
| Thickness/Breadth *(min - max)* | - |

[Submit] [Reset]  Number of records per page  50

Figure 1: Query page of the Dataset

## Query - Results
**Records 1 - 50 of 3556**
Pages:   1 | 2 | 3 | 4 | 5 | 6 | 7 | 8 | 9 | 10 | > >>

**Click on the images or the links below to view more details for each biface.**

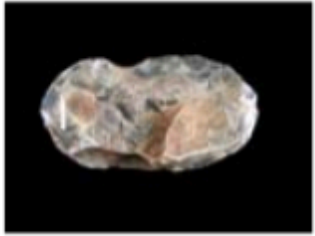

IRREGULAR FLINT HANDAXE (889g)
Location: WARREN HILL ENGLAND
Museum: BRITISH MUSEUM, LONDON, ENGLAND

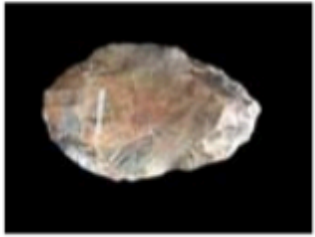

LINEAR FLINT HANDAXE (512g)
Location: WARREN HILL ENGLAND
Museum: BRITISH MUSEUM, LONDON, ENGLAND

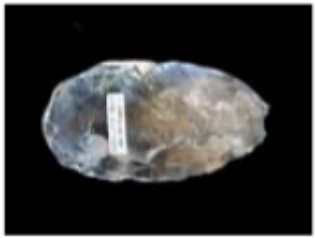

LINEAR FLINT HANDAXE (488g)
Location: WARREN HILL ENGLAND
Museum: BRITISH MUSEUM, LONDON, ENGLAND

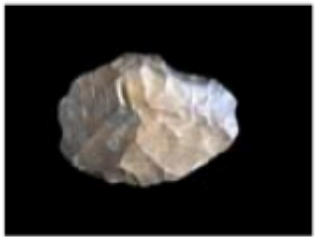

LINEAR FLINT HANDAXE (307g)
Location: WARREN HILL ENGLAND
Museum: BRITISH MUSEUM, LONDON, ENGLAND

Figure 2: Query results view

## Query - Full Record
**Full Record - No. 190**

**Views**

Click on the thumbnails to view and download larger versions:

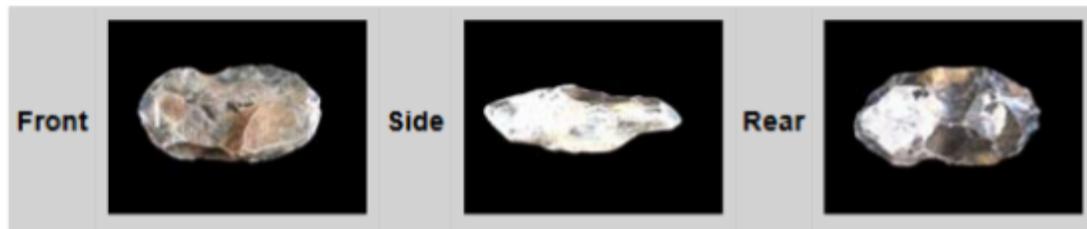

**Description** POSSIBLE ROUGHOUT

**Details**

Explanation of the fieldnames

| Sitename | WARREN HILL |
|---|---|
| Country | ENGLAND |
| Continent | EUROPE |
| Biface type | HANDAXE |
| Completeness | COMPLETE |
| Finder | STURGE |
| Finder's number | UNCLEAR |
| Site subdivision | UNCLEAR |
| Context or level | UNCLEAR |
| Date found | UNCLEAR |
| Museum or holder | BRITISH MUSEUM, LONDON, ENGLAND |
| Museum accession number | 123 |
| Museum accession date | |

Figure 3: Biface record example

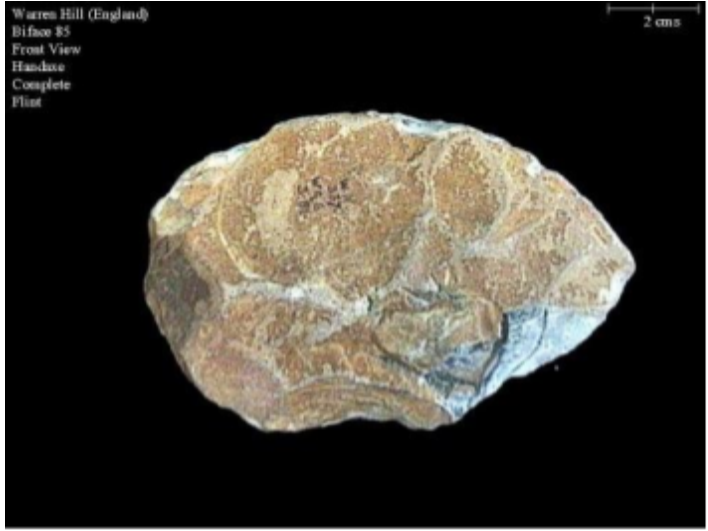

Figure 4: Image 85f.jpg. Front view of biface 85.

There is not a way to bulk download all the records. Instead, the user should open and gather the information of each record independently.

## License

The dataset is licensed under the ADS Terms of Use and Access (*ADS Terms of Use and Access – Archaeology Data Service*, 2025). Under this license, you can:

"*Use materials for research, learning and teaching purposes*

*Adapt, remix, transform and build upon the material for research, learning and teaching purposes*

*Share copy and redistribute the material in any medium or format*". But the license mentions too that *"The data must not be sold or supplied to a third party.", "... duplication or sale of all or part of any of the Data Collections is not permitted, except that material may be duplicated by you for your research use or educational purposes in electronic or print form."* Even when the current project has research and educational purposes, in order to avoid license problems, the original image files will not be shared, and only derived information (scripts, masks, and outlines) is published.

# Web Scraping & Metadata Extraction Pipeline

**Repository**: https://github.com/jfpalomeque/hand_axes_database_downloader

The goal was to design a python script able to check all the records web pages, download the available data and the images, and save everything in a local folder. At the moment of writing this note, urls for the records and images show a stable and repetitive pattern, only changing the record number between them.

The script follows a simple design, using the python library requests to get access to the different urls. To ensure ethical automation, the script development integrated precautions consistent with best-practice guidelines (Paige, 2024). The site's `robots.txt` file was checked before scraping, a randomised delay between requests to avoid overloading the ADS servers, and ADS Terms of Use and Access was checked to confirm that said scrapping would be allowed under them (*ADS Terms of Use and Access – Archaeology Data Service*, 2025).

All the records ids were identified (as explained on the dataset metadata, available on [archaeologydataservice.ac.uk/archives/view/bifaces/help.cfm](archaeologydataservice.ac.uk/archives/view/bifaces/help.cfm)), and for each id, the source code of the record page is analysed, extracting the data of the artifact, and downloading the images. As mentioned, a random delay is added between requests. The data is saved in a tabular file (csv), and a log is generated, to check if the process has failed at any point, allowing to continue from there.

All the script can be run standalone, just cloning the repository ([https://docs.github.com/en/repositories/creating-and-managing-repositories/cloning-a-repository](https://docs.github.com/en/repositories/creating-and-managing-repositories/cloning-a-repository)), and running the following commands (assuming you have Python and pip installed in your machine:

```
pip install -r requirements.txt
python ads_biface_database_scraper.py
```

All images will be downloaded into the *images/* folder, and a *bifaces_records_online.csv* file will be created with all the artifacts details.

## Image Processing & Segmentation Pipeline

**Repository:** [https://github.com/jfpalomeque/hand_axes_image_processing](https://github.com/jfpalomeque/hand_axes_image_processing)

As mentioned earlier, this repository contains

This repository shows a technical demo of an automatic pipeline to prepare and process image collections of archaeological artifacts for computer vision workflows. Specifically, it was developed to work with the Lower Palaeolithic hand-axes / bifaces dataset from the Archaeology Data Service (ADS), when bulk downloading the images and data as explained above..

The primary goals were:

- Convert image filenames to UUID-based names while producing a mapping file.

- Generate a COCO-format JSON for the collection.

- Run an image segmentation pipeline over a collection and embed annotations into the COCO JSON.

The code is lightweight and intended for researchers who want to convert archaeological image collections into formats suitable for machine learning / computer vision experiments

## UUID-based names

An UUID (Universal Unique IDentifier) is an ID that, when generated using the standard procedure, makes the risk of collision (two identical IDs) almost impossible, without the need of central service or server recording and storing all the different IDs.
For this project, the use of UUIDs as the file names allows that collection processed by different systems, teams or researches could be potentially merged without the risk of duplicating said file names. A copy of each image in the collection is saved with said UUID as file name, and the same UUID is saved in the metadata (EXIF) of the image. Original file names and UUIDs are saved in a mapping file, allowing that any additional data about the collection can still be linked with the renamed files.

## COCO-Json format

COCO (Common Objects in Context) is a dataset created by Microsoft (Lin *et al.*, 2014), to show real life examples of objects in their context. In addition to the actual images, a json file was designed, carrying additional information and metadata about the images and objects on each picture, such as location of each object, coordinates of the boundary box, a mask covering the outline of the object, etc...

This provides a standard way to store and share said metadata, allowing the dataset to be used in machine and deep learning applications, such as computer vision segmentation or classification training and test.

In the proposed script, some additional information about the collection is included at the start of the COCO file. Said information is provided in a file called dataset_info.md. The provided information is:

- Description: A short description of the dataset
- Url: Original dataset URL
- Collection_short_name: A short name for the collection. This will be used as file name for the COCO json file
- Version: version of the COCO file
- Year: year of the COCO file
- Contributor: Name and email of the contributor

- Date_created: Date
- Licenses: Licenses under the original dataset was published
    - id: ID of the license
    - name: Name of the license
    - url: URL of the license

For each image in the collection, the UUID, file name and dimensions are recorded in the COCO file.

## Image format and structure

This pipeline assumes that the original images are stored in a folder named `original_images/`. The images should be in standard formats such as JPEG or PNG. The output UUID-named images will be saved in a folder named `uuid_images/`.

The development was done using the mentioned ADS bifaces dataset. In this dataset, images are in JPEG format, with only one artifact per image, typically centered against a plain dark background. Some information and a graphic scale are included in the superior part of the image.

## Segmentation and annotation of the images

In this proposed pipeline, only one object is assumed per image, against a dark background. All possible contours are identified, and the biggest one is considered the artifact in the image. For said object, a binary mask is drawn, and the boundary box coordinates calculated, and saved in the COCO.
A standard category (object) was added to the COCO file, as the current pipeline is not designed to handle multiple categories of objects.

All the script can be run standalone, just cloning the repository (https://docs.github.com/en/repositories/creating-and-managing-repositories/cloning-a-repository), filling in the `dataset_info.md` file with your collection metadata, placing your original images in the `original_images/` folder, and running the following commands (assuming you have Python and pip installed in your machine:

```
pip install -r requirements.txt
python main.py
```

The script will generate the following outputs:

- A COCO-format JSON file with collection metadata, image entries, and segmentation annotations.

- A folder with UUID-named copies of the original images.

- A CSV mapping original filenames to UUID filenames in directory `uuid_images/`.

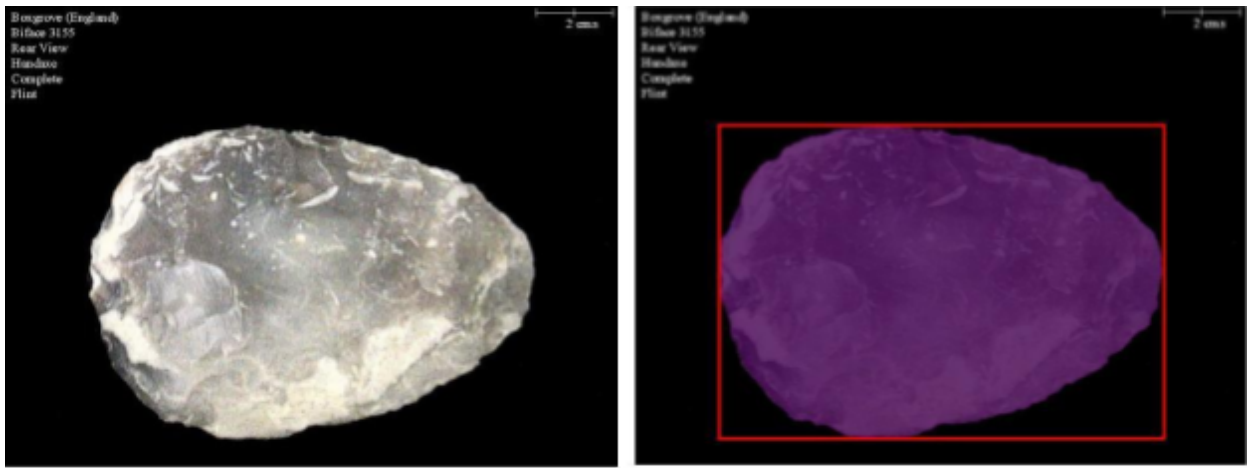

Figure 5: Original image (l) and processed image (r) of rear view, biface 3155, from Boxgrove, England. Segmentation mask in purple, boundaries box in red

## Limitations and future work

The workflow presented here is intentionally narrow and pragmatic, and it comes with several limitations that are important to acknowledge.

First, the image processing pipeline is tailored to a very specific photographic setup: one artifact per image, on a dark background, with labels and a scale bar in a different area of the frame, and with smaller size. Different image formats would require additional techniques, such as background subtraction, learned segmentation models, or even manual annotation.

Second, the segmentation uses classical computer vision methods rather than deep learning. This choice keeps the pipeline lightweight, transparent, and easy to reproduce, but it also means that the performance is sensitive to lighting conditions, object contrast, and noise. Very dark tools on dark backgrounds, strong shadows, or artefacts touching the image border can still cause contour detection errors or incomplete masks.

Third, the scrapping workflow is built around a single case study dataset. Although the code has been written with reuse in mind, other collections will have different metadata structures, file naming conventions, and record layouts. Adapting the scraper to a new ADS collection, or to another repository, will likely require revisiting the HTML parsing logic and configuration files. In addition, changes in the ADS site structure, URL formats, or licensing policy could break parts of the scraper or alter what is allowed. The workflow is therefore best understood as a snapshot of what is possible at the time of writing rather than a permanent interface.

Fourth, the COCO Json format proposed here reflects the needs of the specific dataset used, and may need to be adjusted for other datasets. Additional utilities for adding multiple categories and merging different datasets would need to be developed.

The pipeline could also be generalised to handle multiple objects per image, different background conditions, or alternative photographic conventions. This might involve adding routines for automatic crop proposals, background estimation, or human in the loop correction where confidence is low. On the data side, the COCO extensions could be expanded with additional details.

Finally, although this technical note does not develop a full morphometric analysis, the masks, outlines, and rotated bounding boxes produced here provide an obvious starting point for future studies to explore inter site variation, raw material differences, or comparisons with other lithic datasets.

# Conclusions

This technical note has demonstrated that it is possible to turn a legacy web based archaeological image collection into a structured, segmentation ready dataset using relatively simple tools. By combining a dedicated scraper for the ADS bifaces collection with a classical computer vision pipeline, the workflow moves from point and click access to reproducible, script based retrieval and processing.

The web scraping component provides a way to systematically download images and record level metadata while tracking provenance, respecting repository terms of use, and documenting what was accessed. The image processing pipeline then segments individual artifacts, generates binary masks and bounding boxes, and exports the results in a COCO compatible Json file enriched with archaeologically relevant fields and stable identifiers.

Crucially, the design separates the original images, which remain subject to ADS licensing and are not redistributed, from the derived products that can be shared openly. Scripts, masks, outlines, COCO annotations, and schemas can all be archived and cited, making it easier for others to build upon this work without repeating the most time consuming steps.

Taken together, these components show how existing digital collections can be made more useful for quantitative and computer vision based research without requiring changes to the repositories themselves. The approach is intentionally modest, focusing on a single dataset and a straightforward photographic setup, but it offers a pattern that can be adapted to other collections. By releasing the code and derived data, the project aims to contribute a small but concrete piece of infrastructure to the growing intersection between archaeology, data science, and computer vision.